\documentclass[twocolumn,showpacs,amsmath,amssymb]{revtex4}

\usepackage{bm} 
\usepackage{hyperref} 
\usepackage{natbib} 
\usepackage{subfigure} 
\usepackage{graphicx} 

\newcommand{\sdata}{$N_{\odot,r}$} 
\newcommand{\nggn}{$(n,\gamma)\rightleftarrows(\gamma,n)$}  
\providecommand{\e}[1]{\ensuremath{\times 10^{#1}}}

\renewcommand*{\subfigure}{\alph{subfigure}}

\begin{document}

\title{The Influence of Neutron Capture Rates On The Rare Earth Region Of The r-Process Abundance Pattern}

\author{Matthew R. Mumpower}
\email{mrmumpow@ncsu.edu}
\affiliation{Department of Physics, North Carolina State University, Raleigh, North Carolina 27695-8202,USA}

\author{Gail C. McLaughlin}
\email{gail\_mclaughlin@ncsu.edu}
\affiliation{Department of Physics, North Carolina State University, Raleigh, North Carolina 27695-8202,USA}

\author{Rebecca Surman}
\email{surmanr@union.edu}
\affiliation{Department of Physics and Astronomy, Union College, Schenectady, New York 12308,USA}

\date{\today}

\begin{abstract}
We study the sensitivity of the $r$-process abundance pattern to neutron capture rates along the rare earth region ($A\sim150$ to $A\sim180$). We introduce the concepts of large nuclear flow and flow saturation which determine the neutron capture rates that are influential in setting the rare earth abundances. We illustrate the value of the two concepts by considering high entropy conditions favorable for rare earth peak production and identifying important neutron capture rates among the rare earth isotopes. We also show how these rates influence nuclear flow and specific sections of the abundance pattern.
\end{abstract}

\maketitle

\newpage

\section{Introduction}\label{section1}
The rapid neutron capture process or `$r$-process' has long been known to be an integral component of heavy element nucleosynthesis. The onset of the $r$-process has traditionally been characterized by a relatively large neutron number density ($n_{n}\gtrsim10^{20}$cm$^{-3}$) and high temperature (T$\sim10$GK). These quantities decrease culminating in the last stage of the $r$-process known as \nggn \ freeze-out \cite{Cowan1991}.

During freeze-out nuclides fall out of out of \nggn \ equilibrium and individual neutron capture rates are important in determining final abundances \cite{Beun2009,Surman2009,Arcones2011,Mumpower2011}. For a classical, `hot' freeze-out temperatures around T$\gtrsim1$GK are expected. Recently, Wanajo \cite{Wanajo2007} has suggested that $r$-process nuclides may participate in a `cold' \nggn \ freeze-out, with temperatures as low as (T$\sim.1$GK). Both hot and cold freeze-out scenarios favor short timescales for neutron capture which can first exceed and then compete with $\beta$-decay. Nucleosynthesis in these environments progresses along the NZ-plane, traversing the nuclear landscape including the rare earth region ($A\sim160$) far from the valley of beta stability where little experimental nuclear data exists \cite{Grawe2007}.

The astrophysical location of the $r$-process is not known at this time \cite{Cowan1991,Qian2007,Arnould2007}. There are several candidate sites where the $r$-process may occur. Among the possible candidates are the high entropy senario of supernova ejecta (neutrino driven wind) \cite{Meyer1992,Sumiyoshi2000,Otsuki2000,Thompson2001,Panov2009}, supernova fallback \cite{Fryer2006}, collapse of O-Ne-Mg cores \cite{Wheeler1998,Wanajo2003,Ning2007}, gamma ray burst accretion disks \cite{Surman2004,McLaughlin2005,Surman2006}, and various compact object merging scenarios \cite{Freiburghaus1999,Goriely2005,Surman2008}. Observational data from metal-poor stars \cite{Sneden2000,Sneden2003,Ivans2006} favors massive stars which mature on short timescales. A comparison of sites has suggested that core-collapse supernova should be favored over neutron star mergers for the production of the heaviest $r$-process elements \cite{Qian2005}. However, modern supernova calculations typically do not achieve sufficiently neutron rich conditions conducive to heavy element nucleosynthesis \cite{Arcones2007,Hudepohl2010,Fischer2010}.

\begin{figure*}[htp]
      \includegraphics[width=85mm,height=70mm]{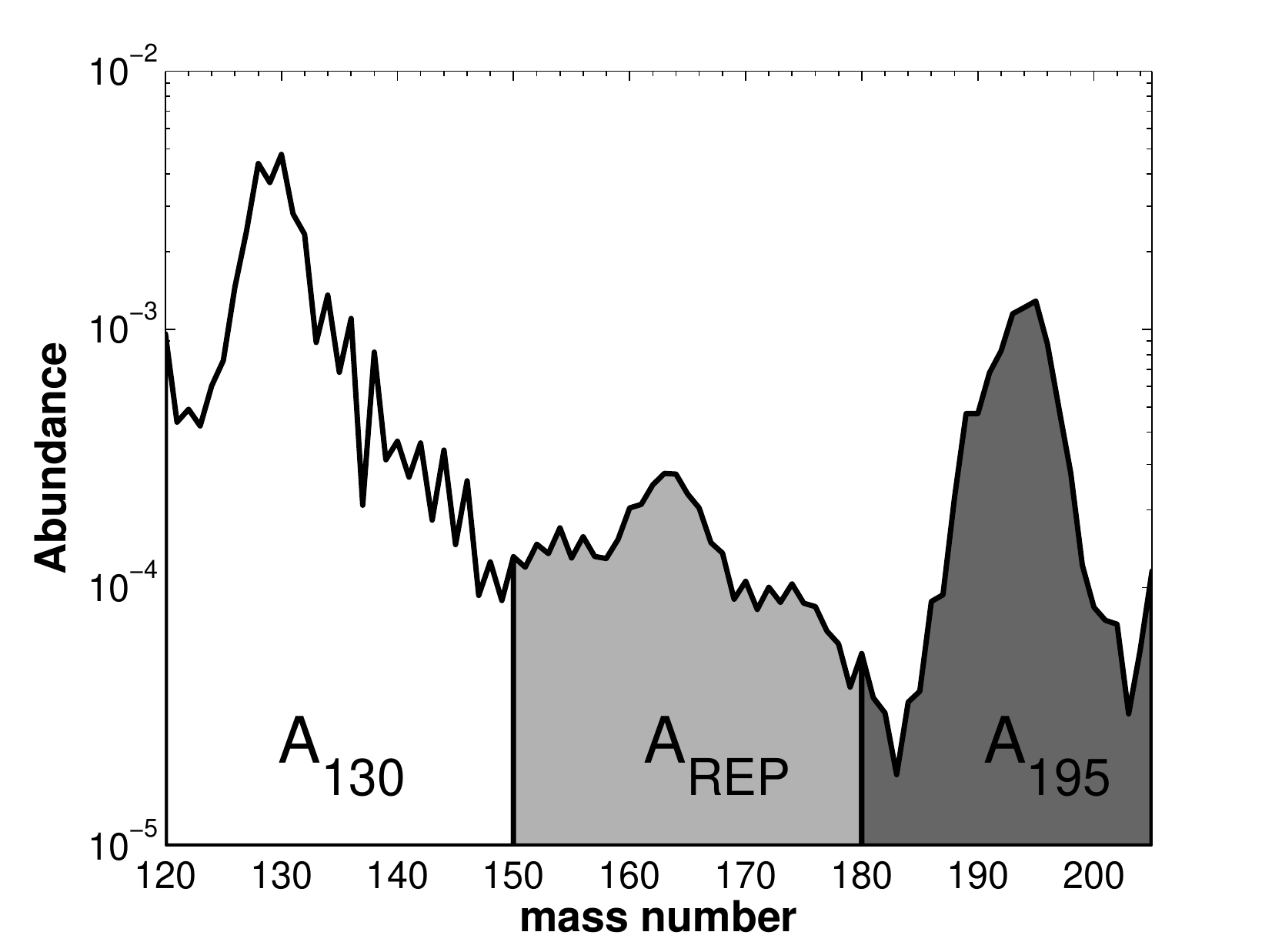}
      \caption{\label{fig:sdata} Shows the solar $r$-process abundance pattern, \sdata \ versus atomic mass (data from \cite{Kappeler1989}). The three main regions of the pattern are highlighted. The A=130 peak is shaded white, the rare earth elements are shaded light gray and the A=195 region is shaded in dark gray. Abundance scale is arbitrary.}
\end{figure*}

Our understanding of the $r$-process is also naturally entangled with our knowledge of the input nuclear physics. For example, it is well known that the long $\beta$-decay half-lifes found at closed neutron shells are responsible for the $A=130$ and $A=195$ peaks found in the $r$-process abundance pattern \cite{BBFH1957}, see Figure \ref{fig:sdata}. While it is difficult to measure the properties of short-lived nuclides far from stability there have been many recent advances \cite{Hosmer2005,Matos2008,Rahaman2008,Baruah2008,Jones2009,Hosmer2010,Nishimura2011,Quinn2011}. Future radioactive beam facilities are expected to help in this endeavor. Due to limited data on nuclides far from stability, theoretical extrapolations must be used as input for $r$-process calculations.

The primary nuclear physics inputs needed to determine the nucleosynthetic outcome include neutron capture cross sections, separation energies, masses, $\beta$-decay rates, and branching ratios. The effects of different masses \cite{Pfeiffer1997} and $\beta$-decay rates \cite{Borzov1997,Montes2006} have been studied in detail for some time. Until recently however, neutron capture rate cross sections have warranted less consideration.

In the past decade studies of neutron capture rates have been performed by several groups \cite{Goriely1997,Goriely1998,Rauscher2005,Farouqi2006,Surman2009,Beun2009}. Two recent studies of particular interest have been performed by Beun et al. \cite{Beun2009}, and Surman et al. \cite{Surman2009} who demonstrated capture rate effects of \textit{individual} isotopes. The former's study focused on the single neutron capture of $^{130}\text{Sn}$. The effects of elements near the $A=130$ peak were studied in detail by Surman et al. \cite{Surman2009}. In each of these two studies the neutron capture rates of \textit{single} elements in the $A=130$ region lead to \textit{global} changes in the final $r$-process abundance pattern.

It has been shown in \cite{Mumpower2011,Surman1997} that neutron capture rates are important for correctly forming the rare earth peak. In order to understand observational data e.g. \cite{Sneden2008,Lodders2009}, it is crucial to determine which rates are important and how these rates influence nuclear flow and ultimately final abundances. If in the future improved calculations or measurements are included this will increase the efficacy of the rare earth peak as a freeze-out diagnostic \cite{Mumpower2012}.

In this paper we study neutron capture rates in the rare earth region using ideal freeze-out conditions for this sector \cite{Mumpower2011,Mumpower2012}. Modifying individual neutron capture rates of nuclides in this region, we identify the most influential and in return show their effects on sections of the $r$-process abundance pattern. We also isolate factors which influence the magnitude of this neutron capture rate effect.

\section{Calculations}\label{section2}
To study the effects of neutron capture rates on final abundances we implement a one-dimensional model of the $r$-process by following the abundance composition of a single ejected mass zone. It was shown in \cite{Mumpower2011,Mumpower2012} that the rare earth elements are sensitive to the thermodynamic evolution of the ejected material. Thus, to simulate the qualitative behavior of hydrodynamic outflows we parameterize our trajectories via two analytic procedures. For the hot freeze-out evolution we parameterize the density with the same functional form as \cite{Meyer2002},

\begin{equation}\label{eqn:bradRho}
\rho(t)=\rho_{1}\text{exp}(-t/\tau)+\rho_{2}\left(\frac{\Delta}{\Delta+t}\right)^{2}
\end{equation}
where $\rho_{1}+\rho_{2}$ is the density at time $t=0$, $3\tau=\tau_{dyn}$ and $\Delta$ chosen so that the two terms on the right hand side are equivalent at a time $t=\tau$. The first term controls the trajectory at early times, during which the neutron-to-seed ratio is set and the second term controls the late time behavior, during which the rare earth peak forms.

For cold freeze-out evolutions \cite{Wanajo2007} we use a second density parameterization by Panov and Janka \cite{Panov2009},
\begin{subequations}
 \label{eqn:pjRho}
 \begin{eqnarray}
\rho(t)=\rho_{0}\text{exp}(-t/\tau) \label{eqn:pjRhoA} \\
\rho(t)=\rho_{0}\left(\frac{t}{t_{0}}\right)^{-2} \label{eqn:pjRhoB}
 \end{eqnarray}
\end{subequations}
where $\rho_{0}$ is the density at time $t=0$ and $t_{0}$ is the switch over point corresponding to a temperature of $T_{9}=2$. Equation \ref{eqn:pjRhoA} gives early time behavior of the outflow and equation \ref{eqn:pjRhoB} pertains to late time behavior.

In modern $r$-process studies it is typical to approximate the thermodynamics by assuming the ejecta is radiation dominated. However, we require the full thermodynamics and so we instead use the solver from \cite{McLaughlin1999} to generate all trajectories.

Our nucleosynthesis calculations consist of a multi-tiered algorithm which is coupled by three canonical stages. This nucleosynthesis network code was used in previous neutron capture studies by both Beun \cite{Beun2009} and Surman \cite{Surman2009}.

The simulation starts with the regime of nuclear statistical equilibrium. During this stage entropy and density effectively determine all thermodynamic quantities. The second stage of the simulation employs an intermediate reaction network \cite{Hix1999} with PARDISO solver \cite{Schenk2004,Schenk2006}. At this point the calculations include electromagnetic, strong and weak interactions.

The third and final stage of our network calculation consists of a reduced $r$-process network as described in \cite{Surman1997,Surman2001}. The primary reaction channels for nuclides in this section of the reaction network are beta-decay, neutron capture, and photo-dissociation. A detailed analysis regarding the formation of the rare earth elements during this stage is given in \cite{Mumpower2011,Mumpower2012}.

In the third stage we track individual nuclear abundances by solving a set of differential equations given by the short hand notation (see \cite{Clayton1968} and \cite{Cowan1991} for details),
\begin{equation}\label{ydot}
\dot{Y}(Z,A)=\sum_{Z',A'}\lambda_{Z',A'}Y_{Z',A'}+\sum_{Z',A'}\rho N_{A}\langle\sigma v\rangle_{Z',A'}Y_{Z',A'}Y_n
\end{equation}
where the quantity $\dot{Y}(Z,A)$ represents the time rate of change of abundance for nuclide $(Z,A)$, $Y_{Z',A'}$ is the abundance for nuclide $(Z',A')$, $\rho$ is the density, $N_{A}$ is Avagardo's number, $\langle\sigma v\rangle$ is Maxwellian averaged neutron capture cross section for nuclide $(Z',A')$ and $Y_n$ the free neutron abundance. The first term on the right hand side holds information about beta decay modes and photodisintegrations. The rate, $\lambda_{Z',A'}$ could be one of following: the $\beta$-decay rate with emission of $j$ neutrons, $\lambda_{\beta jn}$ or the photo-dissociation rate, $\lambda_{\gamma}$. The second term on the right hand side includes reactions with neutrons. 

Nuclear flow is monitored by following individual terms in equation \ref{ydot}. For example, in the following equation we explicitly write out each term which gives the total flow in and out of a nucleus:
\begin{subequations}
 \label{ydot1}
 \begin{eqnarray}
\dot{Y}_{in}(Z,A)=\lambda_{\beta}(Z-1,A)Y(Z-1,A)& \nonumber \\ 
  +\lambda_{\beta 1n}(Z-1,A+1)Y(Z-1,A+1) \nonumber \\
  +\lambda_{\beta 2n}(Z-1,A+2)Y(Z-1,A+2) \nonumber \\
  +\lambda_{\beta 3n}(Z-1,A+3)Y(Z-1,A+3) \nonumber \\
  +\lambda_{\gamma}(Z,A+1)Y(Z,A+1) \nonumber \\ 
  +\langle\sigma v\rangle_{Z,A-1}Y(Z,A-1)\rho N_{A}Y_n\label{ydot1a} \\
 \nonumber \\
\dot{Y}_{out}(Z,A)=[\lambda_{\beta}(Z,A)+\lambda_{\beta n}(Z,A) \nonumber \\
  +\lambda_{\beta 2n}(Z,A)+\lambda_{\beta 3n}(Z,A)]Y(Z,A) \nonumber \\
  +\lambda_{\gamma}(Z,A)Y(Z,A)+\langle\sigma v\rangle_{Z,A}Y(Z,A)\rho N_{A}Y_n\label{ydot1b}
 \end{eqnarray}
\end{subequations}
The difference of the two terms in equation \ref{ydot1} yields equation \ref{ydot}.

In general neutron reaction rates are of the following functional form (Fowler et al \cite{Fowler1967}):
\begin{equation}\label{eqn:ncr}
\langle\sigma v\rangle=\left(\frac{8}{\mu\pi}\right)^{1/2}(kT)^{-3/2}\int_{0}^{\infty}E\sigma(E)\exp(-E/kT)dE
\end{equation}

Neutron capture cross sections can vary by orders of magnitude between nuclear models \cite{Beun2009} so we explore a variety of nuclear physics input: Finite Range Droplet Model (FRDM) \cite{Moller1995,ADANDT2000,NONSMOKER1998}, Extended Thomas-Fermi with Strutinsky Integral and Quenching (ETFSI) \cite{ETFSIQ1996,ADANDT2000,NONSMOKER1998} version 17 of the Hartree Fock Bogoliubov model (HFB17) \cite{HFB17}\footnote{http://www.astro.ulb.ac.be/} and Duflo-Zuker (DZ) \cite{DZ}. The neutron capture rates of HFB17 and DZ were calculated using the TALYS code \cite{TALYS}. The $\beta$-decay rates used in our $r$-process network come from \cite{Moller2003}. Calculations with different nuclear models show qualitatively similar behavior. So in our analysis we focus the discussion using one nuclear model, FRDM. We display results for all three nuclear models in section \ref{sec:npu}. 

For the conditions we use entropy per baryon in units of Boltzmann's constant, $S=165$, dynamical timescale, $\tau_{dyn}=85$ms and electron fraction, $Y_{e}=.30$ for a hot freeze-out and $S=105$, $\tau_{dyn}=50$ms and $Y_{e}=.30$ for a cold freeze-out. These conditions were chosen based on the procedure given in \cite{Mumpower2010}.

\section{Results}\label{results}
\subsection{Neutron Capture Rate Studies}
A neutron capture rate study consists of a `baseline' simulation where the appropriate input of astrophysical and nuclear parameters are established. The final output is an abundance pattern ($Y^{baseline}_{A}$) as a function of atomic mass, $A$. Subsequent simulations are then conducted with the \textit{same input data} but with the neutron capture rate of a \textit{single} isotope in the rare earth peak changed by a factor of $K=5$, $10$, $50$, or $100$; each producing a final abundance pattern, $Y^{K}(A)$.

A change in the neutron capture rate can lead to one of two distinct physical processes. A `neutron capture effect' occurs when change in capture rate results in pathway changes that shift material from the nucleus $(Z,A)$ whose rate has been changed, to a nucleus $(Z,A+1)$, i.e. to the right on the NZ-plane. Or a `photo-dissociation effect' which occurs when a change in capture rate on nucleus $(Z,A)$ results in pathway changes that shift material from nucleus $(Z,A+1)$ to the nucleus $(Z,A)$, i.e. to the left on the NZ-plane. We only find a neutron capture effect among the rare earth elements as photo-dissociation effects typically require highly populated nuclei, see \cite{Surman2009} for a detailed discussion of this effect.

For each study, we consider the magnitude of influence a single neutron capture rate has on the \textit{final} abundance pattern using the quantity $F$.
\begin{equation}\label{eqn:F}
F_{K}=100\sum_{A}\frac{|Y^{K}(A)-Y_{A}^{baseline}|}{Y_{A}^{baseline}}
\end{equation}

A larger $F$ value represents more deviation (percent change) from the baseline simulation. If $F=0$ then the neutron capture simulation abundance pattern and the baseline abundance pattern are equal; the neutron capture rate change had no effect on the abundance pattern. 

\begin{figure*}[htp]
      \includegraphics[width=175mm,height=200mm]{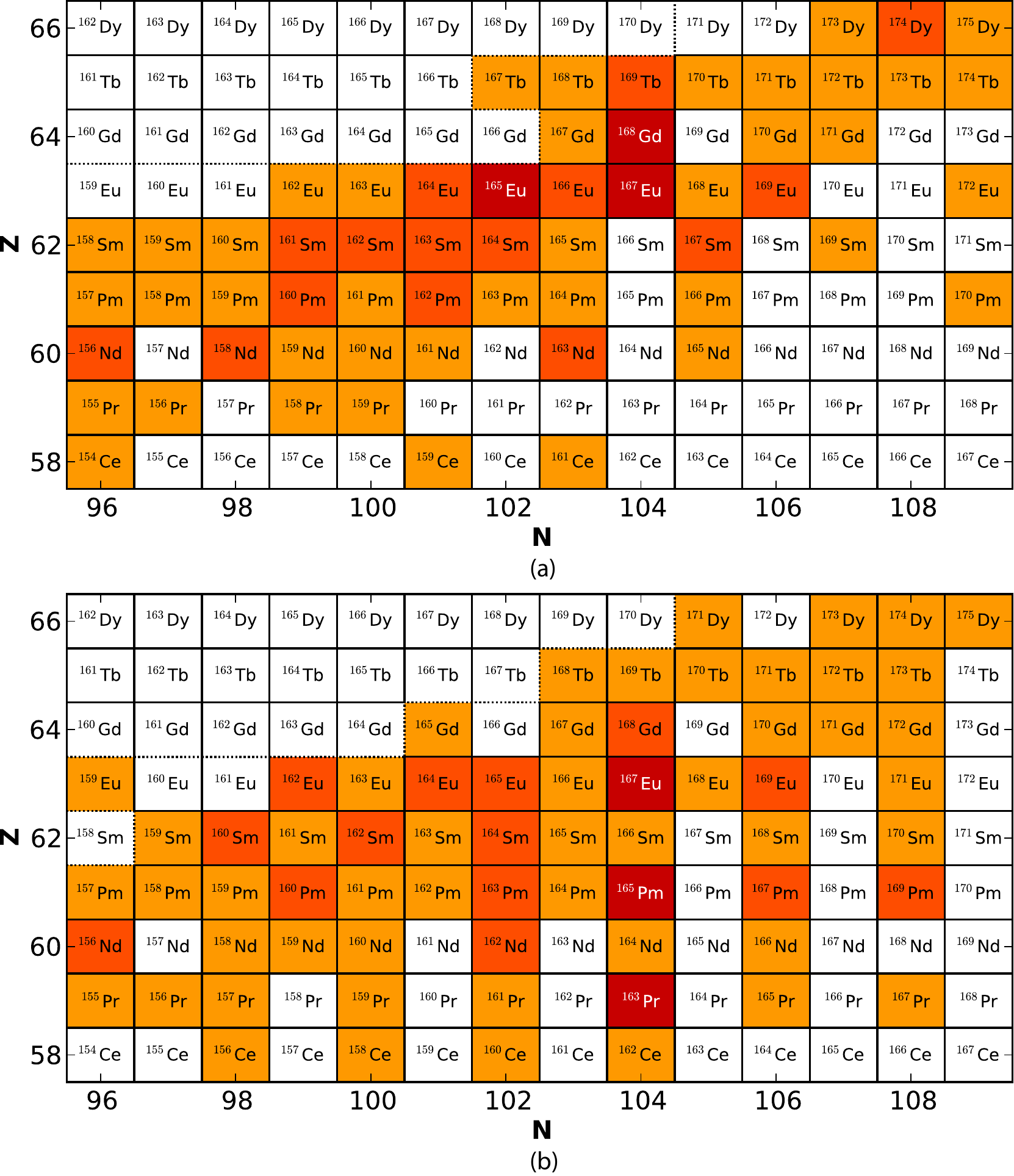}
      \caption{\label{fig:fgrid} Shows neutron capture rates that significantly influence the abundance pattern. The results of two separate neutron capture rate studies are shown for a hot freeze-out, panel (a) and cold freeze-out, panel (b). In both cases individual neutron capture rates were changed by $K=10$. Darker shades represent an increased effect on the abundance pattern. In order of lightest to darkest each shade represents: white ($0\leq F\leq100$), light ($100<F\leq200$), medium ($200<F\leq400$), darkest ($F>400$). Above the dotted line neutron capture flows are not large enough for an increase in a neutron capture rate to produce a significant neutron capture effect.}
\end{figure*}

We highlight the results of two different neutron capture rate studies containing 126 nuclei in Figure \ref{fig:fgrid}. Each element's neutron capture rate has been changed by a factor of $K=10$ and the magnitude of its effect on the final abundances abundance is denoted by the degree of shading. A darker gray (darker red online) represents a larger effect. White boxes represent capture rates with little to no effect on the final abundance pattern ($0\leq F\leq100$). Each darker shaded nuclei represents an approximate factor of two increase in the $F$-measure with the darkest shade representing $F\gtrsim400$. Panel (a) of Figure \ref{fig:fgrid} shows the effects of neutron capture rate changes under a hot evolution and panel (b) shows the effects of neutron capture rate changes under a cold evolution.

\subsection{Comparing Hot and Cold Evolutions}
While the overall distribution of influential nuclei in Figure \ref{fig:fgrid} is similar between the hot and cold freeze-out trajectories there are two prominent differences between these environments. First, there is a visible favoring of nuclei with even number of neutrons (even-N effect) in the cold evolution and an odd-N effect occurring at early times in the hot freeze-out evolution. Second, the magnitude of the neutron capture rate effect can vary for the same nuclei under the two $r$-process environments throughout the central shaded region in Figure \ref{fig:fgrid}.

The $r$-process path is the time sequenced set of most abundant isotopes. It is useful to follow the path through the NZ-plane to understand the distribution of influential nuclei across the rare earth region. In Figure \ref{fig:fgrid}, the path begins far from stability in the lower right corner of the NZ-plane progressively making its way back towards stability (top left) as the temperature falls and $\beta$-decay begins to dominate the nuclear flow.

As the path moves through the bottom right corner of the hot evolution \nggn \ equilibrium is still in effect. Under these conditions small changes in neutron capture rates have no effect on the flow of material through the particular isotope. Due to equilibrium, the flow simply readjusts to compensate for the change. In the cold evolution, all photodisintegrations have frozen out and so the path is controlled by neutron captures and $\beta$-decays only \cite{Arcones2011,Mumpower2011}. In this case, changes to neutron capture rates can impact final abundances.

The path next encounters slower neutron capture rates in the central region of the figure. For those nuclei which are out of equilibrium and have significant abundance, changes in capture rate can now produce measurable effects \cite{Mumpower2011}.

The nuclei in the top left portion are populated primarily via $\beta$-decay. All other reactions, including neutron capture have frozen out. Thus we find little to no impact of neutron capture rates in this region.

\subsection{Changes to the Final Abundance Pattern}
\begin{figure*}[htp]
      \includegraphics[width=160mm,height=60mm]{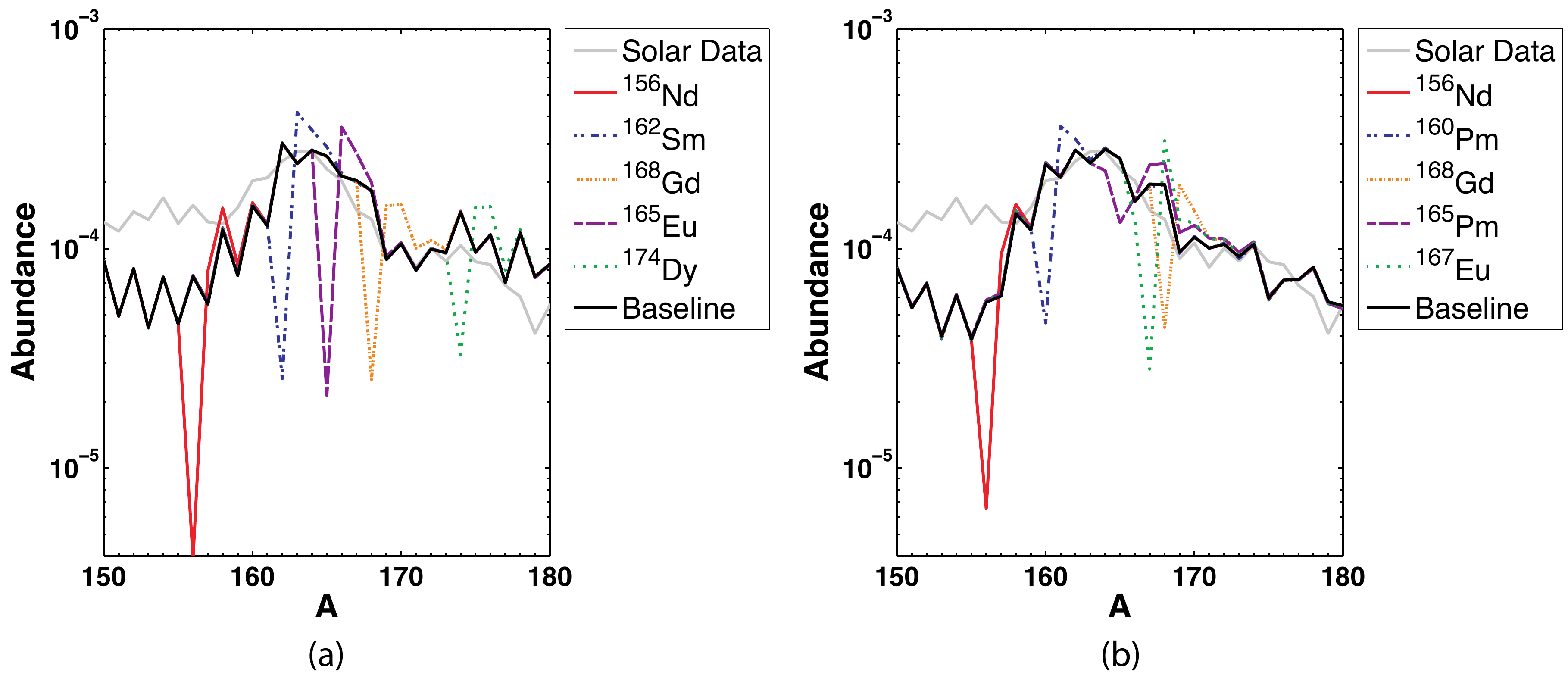}
      \caption{\label{fig:abncr} Shows the effect of particular neutron capture rates on the rare earth abundances. Simulations were performed using hot (panel a) cold (panel b) freeze-out evolutions. The baseline curve, $Y^{baseline}$ is represented by a bold black line and the solar data is repressented by a solid gray line. For both types of trajectories we show five curves, $Y^{K=10}$ each representing a simulation where a single neutron capture rate was changed by a factor of $10$.}
\end{figure*}

We highlight the effect of an individual neutron capture rate change has on the final abundances in Figure \ref{fig:abncr}. These nuclides were chosen from the hot and cold neutron capture studies of Figure \ref{fig:fgrid}. Each neutron capture rate has been changed by a factor of $K=10$ and the resultant final abundance pattern is compared to both the baseline and solar abundances.

Inspection of Figure \ref{fig:abncr} shows that changes to neutron capture rates in the rare earth region produce only local changes in final abundances and that these changes are significant even for changes by a factor of $K=10$. In both hot and cold evolutions $^{156}\text{Nd}$ shows a similar behavior under change in capture rate. This is not always true for each nuclide, compare e.g. $^{168}\text{Gd}$. We also find that individual nuclei exhaust their neutron capture effect at different values of $K$. For instance, in the hot environment, the $^{165}\text{Eu}$ neutron capture effect is exhausted near $K=50$ while in the cold environment, the $^{165}\text{Pm}$ neutron capture effect is maximal near $K=10$.

\section{Analysis Of The Neutron Capture Effect}\label{sec:mnce}

\begin{figure*}[htp]
      \includegraphics[width=170mm,height=80mm]{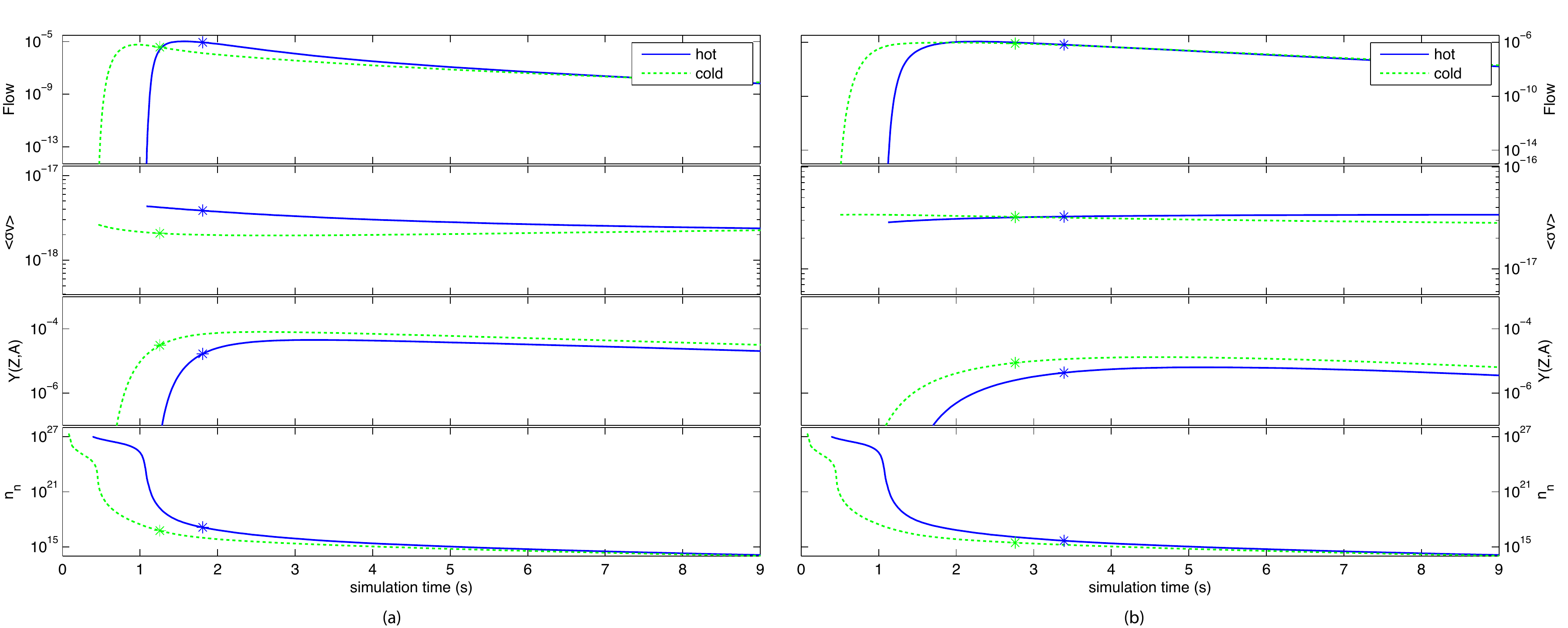}
      \caption{\label{fig:fcnc} Compares neutron capture flows in baseline simulations for the hot (solid) and cold (dashed) evolutions as a function of simulation time. Also shown are the neutron capture rate, nuclear abundance and neutron number density. The product of these three determines the neutron capture flow. The neutron capture rate change has the most influence around the time when $\dot{S}$ is maximal (star). The left panel highlights $^{168}$Gd and the right panel $^{171}$Dy.}
\end{figure*}

We now consider the factors which influence the magnitude of the neutron capture effect. A necessary condition for large neutron capture effect is that the nuclide exhibit large nuclear flow through the neutron capture channel. In order for a large flow through the neutron capture channel to have a significant influence on the abundance pattern the neutron capture flow must also be unsaturated. Flow saturation, which we denote as $\dot{S}$, occurs when the sum of material flowing through input channels matches the flow of a single output channel. In the case where the output channel is neutron capture the channel is `saturated' ($\dot{S}=0$) when it is matched by photo-dissociation and $\beta$-decay in-flows. We use these concepts to explain the relative differences seen in $F$ between nuclides in the hot and cold evolutions.

\subsection{Large Nuclear Flow}
A large neutron capture flow means significant material transportation through the neutron capture reaction channel. The last summand in equation \ref{ydot1b} contains the relevant information on the movement of material via neutron capture flow out from isotope $(Z,A)$ to isotope $(Z,A+1)$. We provide it for convenience:
\begin{equation}\label{eqn:ncflow}
\text{neutron capture flow}=\langle\sigma v\rangle_{Z,A}Y(Z,A)\rho N_{A}Y_n
\end{equation}
This equation consists of three main ingredients: the thermally averaged neutron capture cross section, $\langle\sigma v\rangle_{Z,A}$, the abundance of the particular isotope, $Y(Z,A)$ and the neutron number density, $\rho N_{A}Y_n$. The interplay between these three components determines the size of the neutron capture flow.

For the energy ranges explored in the two astrophysical environments considered here, $\langle\sigma v\rangle_{Z,A}$ is a relatively flat function of temperature. Therefore, for a fixed nuclear data set the differences between classical and cold neutron capture flows is primarily due to differences in the neutron number density and elemental abundance. Both components are significantly influenced by the astrophysical environment and nuclear data set. Figure \ref{fig:fcnc} shows the interplay between the three components of equation \ref{eqn:ncflow} as a function of simulation time for the baseline case. Two nuclides are highlighted, $^{168}$Gd in the left panel and $^{171}$Dy in the right panel.

To maximize the neutron capture rate effect we search for large out of equilibrium neutron capture flows in the baseline simulation. In order to measure the magnitude of the neutron capture flow in the baseline simulation we compute,
\begin{equation}\label{eqn:ncflowRatio}
L=\frac{\int{\langle\sigma v\rangle_{Z,A}Y(Z,A)\rho N_{A}Y_n}dt}{Y^{\ast}(Z,A)}
\end{equation}
where $Y^{\ast}(Z,A)$ is the abundance when the nucleus is furthest from saturation ($\dot{S}$ maximal), the integrand is equation \ref{eqn:ncflow} and the integral is taken over simulation time.

When $L\gtrsim.2$, flow through the neutron capture channel is sufficiently large for a change in neutron capture rate to produce significant change in the final abundances. We highlight this in both panels of Figure \ref{fig:fgrid} by a dotted line. Above the dotted line, $L<.2$ and below the dotted line $L>.2$. The measure $L$ shows quantitatively how changes in the neutron capture rates of nuclei above the dotted line have no influence on the final abundances.

\subsection{Flow Saturation}
In the region of Figure \ref{fig:fgrid} where $L>.2$, we can use flow saturation to understand the differences in the magnitude of the neutron capture effect among nuclei.

To measure saturation in the neutron capture channel we take equation \ref{ydot1a} and subtract equation \ref{eqn:ncflow},
\begin{equation}\label{eqn:sat}
\dot{S}=\dot{Y}_{in}(Z,A)-\langle\sigma v\rangle_{Z,A}Y(Z,A)\rho N_{A}Y_n
\end{equation}
Saturation occurs when $\dot{S}=0$. Changes to a neutron capture rate under saturation have no effect on the flow of material because the output channel is limited by the in-flowing channels. When $\dot{S}<0$, more material is flowing out through the neutron capture channel than is flowing into the input channels. When $\dot{S}>0$, the neutron capture flow is smaller than the in-flowing channels. 

For a large neutron capture effect it is crucial that $\dot{S}$ is large and greater than zero so that the neutron capture flow is furthest from saturation. We can approximate the time at which a large neutron capture flow is important in producing a large neutron capture effect in each simulation by finding the maximum of $\dot{S}$.

When predicting the magnitude of the neutron capture effect it is useful to define the integral of equation \ref{eqn:sat} over simulation time for $\dot{S}>0$,
\begin{equation}\label{eqn:satint}
S=\int\limits_{\dot{S}>0} \! \dot{S} \, dt
\end{equation}
which we call the unsaturated index. Physically, it is the amount of material that is flowing into the nucleus but is not flowing out via neutron capture. Thus, this material could be directed out through the neutron capture channel if the neutron capture rate were increased.

A special case of neutron capture flow saturation occurs as nuclei fall out of \nggn \ equilibrium in hot scenarios. If the temperature is high enough photo-dissociation rates are large so that the net flow to the right can be limited by leftward flowing material in the photo-dissociation channel. If the nuclide is in \nggn \ equilibrium then the photo-dissociation and neutron capture terms in equation \ref{eqn:sat} are large and cancel so that $\dot{S}\approx0$, see bottom right corner of panel (a) of Figure \ref{fig:fgrid}.

Odd-N nuclei are particularly susceptible to flow saturation as they have smaller separation energies than even-N nuclei. This means that for odd-N nuclei the neutron capture photo-dissociation rate pair $(Z,A)$ and $(Z,A+1)$ tends to fall out of equilibrium sooner than the rate pair $(Z,A-1)$ and $(Z,A)$ \cite{Surman2009}. Thus, the $F$-measure is sensitive to neutron capture rates on odd-N nuclei far from stability in a hot freeze-out. For example, in the top panel of Figure \ref{fig:fgrid}, $^{161}\text{Ce}$,$^{165}\text{Nd}$, $^{166}\text{Pm}$ and $^{170}\text{Pm}$ all fall out of equilibrium earlier than the surrounding nuclei and each exhibit a small neutron capture effect limited by flow saturation. 

Flow saturation also occurs in cold environments. In this case, neutron capture rate effects are limited for odd-N rare earth isotopes because they have faster neutron capture rates and $\beta$-decay rates compared to even-N rare earth isotopes. Faster neutron capture rates in odd-N nuclei means the first term in equation \ref{eqn:sat} is larger for even-N nuclei than for odd-N nuclei. Faster beta decay rates in odd-N nuclei also implies the first term in equation \ref{eqn:sat} is larger for even-N nuclei than for odd-N nuclei. The net effect is that neutron capture flows of odd-N nuclei are closer to saturation ($\dot{S}\approx0$) than the flows of even-N nuclei. Therefore, even-N nuclei are favored by the $F$-measure, see bottom right corner of panel (b) of Figure \ref{fig:fgrid}.

\subsection{Flow Saturation As A Predictor For The Magnitude Of The Neutron Capture Effect}
Computing the unsaturated index using the baseline simulations, we can now predict and understand the relative differences observed in the $F$-measure of the same nuclei in the central region of Figure \ref{fig:fgrid} between hot and cold freeze-out conditions.

For $^{168}$Gd the unsaturated index is larger in the hot baseline, $S=6.73\e{-3}$ than in the cold baseline, $S=1.27\e{-3}$. This implies the neutron capture effect should be larger in the hot scenario as can be verified by comparing the two panels in Figure \ref{fig:fgrid}. Saturation also estimates when the neutron capture effect is important, $\dot{S}$ maximal. Returning to Figure \ref{fig:fcnc}, we see that this point (star on the figure) is sensitive to thermodynamic conditions and may occur when the neutron capture flow is not at its largest value.

The neutron capture flow components of $^{171}$Dy near the time of maximal $\dot{S}$ also leads to a large neutron capture flow, albeit a magnitude smaller than the neutron capture flow of $^{168}$Gd. For $^{171}$Dy the unsaturated index is larger in the cold baseline, $S=2.44\e{-4}$ as compared to the hot baseline, $S=1.23\e{-4}$. This implies the neutron capture effect should be larger in the cold senairo and can again be verified by comparing the two panels in Figure \ref{fig:fgrid}.

Explaining the relative differences seen in $F$ by comparing unsaturated indices and then confirming the result with $F$-measure values works extremely well. However, a word of caution is necessary: The $F$ defined in equation \ref{eqn:F} is a sum of \textit{percent} abundance differences and the unsaturated index, equation \ref{eqn:satint}, contains \textit{total} abundance yield information. Thus if one prefers exact agreement between the result of the capture study ($F$-measure value) and the unsaturated index, one should only use \textit{differences} in abundance or massfraction as the $F$-measure (for example, see $F$ defined in \cite{Surman2009}).

\section{Analysis: Pathway Changes}
In the previous sections we have discussed the magnitude of the neutron capture rate effect and demonstrated how changes in neutron capture rates can influence the final abundance distribution. We now turn our attention to the pathway changes produced by the rate changes, focusing on individual nuclei in the $K=10$ case.

In order to study changes in the path, for each element in the network we produce a real number $\Delta Y(Z,A)$ representing the total change of the abundance yield over $r$-process simulation time,
\begin{equation}\label{deltadef}
\Delta Y(Z,A)=\int{[Y_{Z,A}^{ncr}(t)-Y_{Z,A}^{baseline}(t)]}dt
\end{equation}
where $Y_{Z,A}^{ncr}$ is the abundance of isotope $(Z,A)$ when one neutron capture rate has been changed and $Y_{Z,A}^{baseline}$ is the abundance of isotope $(Z,A)$ in the baseline simulation. If $\Delta Y(Z,A)$ is positive then more material resides in the particular nucleus during the capture rate simulation than during the baseline simulation. Conversely, less material resides in the nucleus during the capture rate simulation than during the baseline simulation if the value of $\Delta Y(Z,A)$ is negative. For capture rate changes in the rare earth region most nuclei in the network have $\Delta Y(Z,A)=0$, except for locally around the nucleus whose capture rate has been changed.

\begin{figure*}[htp]
      \includegraphics[width=140mm,height=60mm]{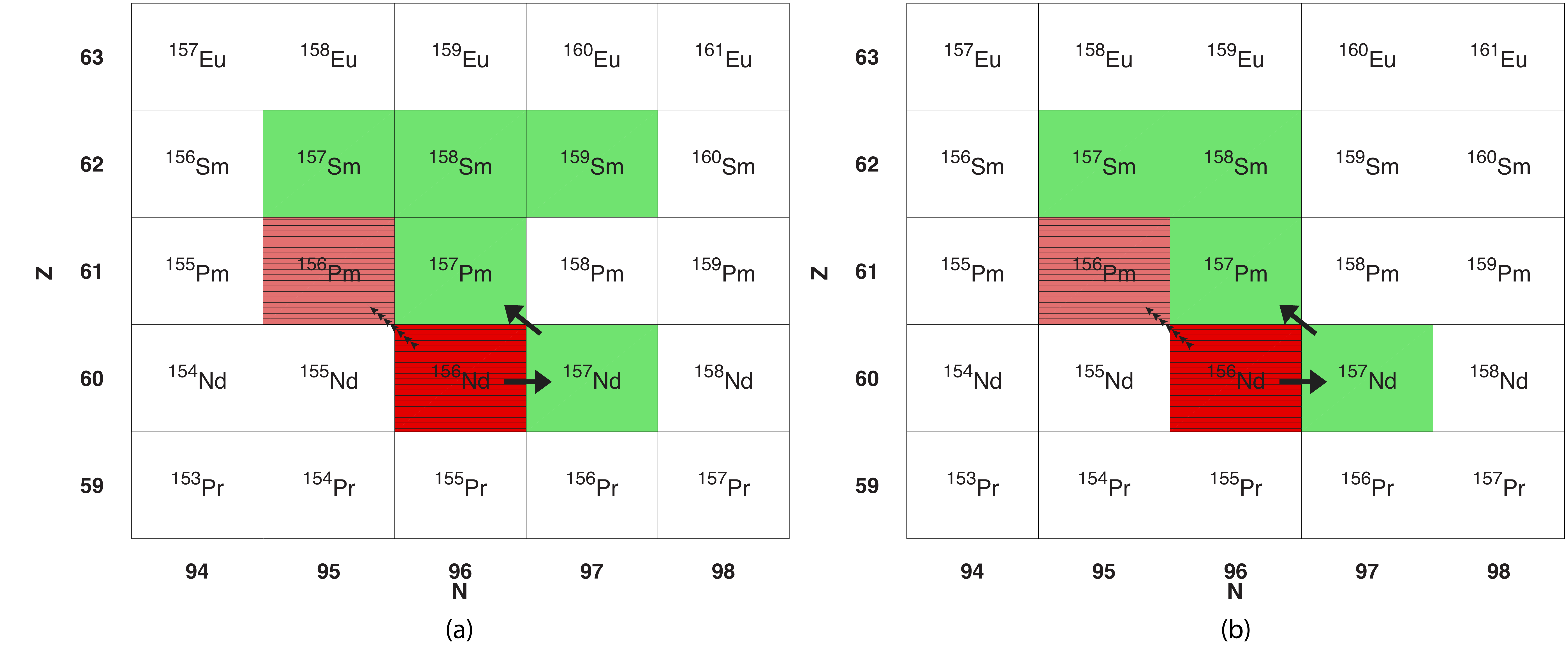}
      \caption{\label{fig:path1} Shows the change in the nucleosynthetic pathway when the neutron capture rate of $^{156}\text{Nd}$ is changed. In the baseline simulation the flow of material through $^{156}\text{Nd}$ occurs primarily in the $\beta$-decay channels. When the neutron capture rate is increased by a factor of $K=10$, the flow out of $^{156}\text{Nd}$ is primarily through neutron capture. This occurs in both types of trajectories hot (left panel) and cold (right panel). Jagged arrows represent nuclear flow in the baseline simulation while solid bold arrows represent flow with the changed capture rate. Relative decreases in abundances along the path are denoted by hatched gray tones (online red) while relative increases are represented by solid gray tones (online green).}
\end{figure*}

\begin{figure*}[htp]
      \subfigure{\includegraphics[width=140mm,height=60mm]{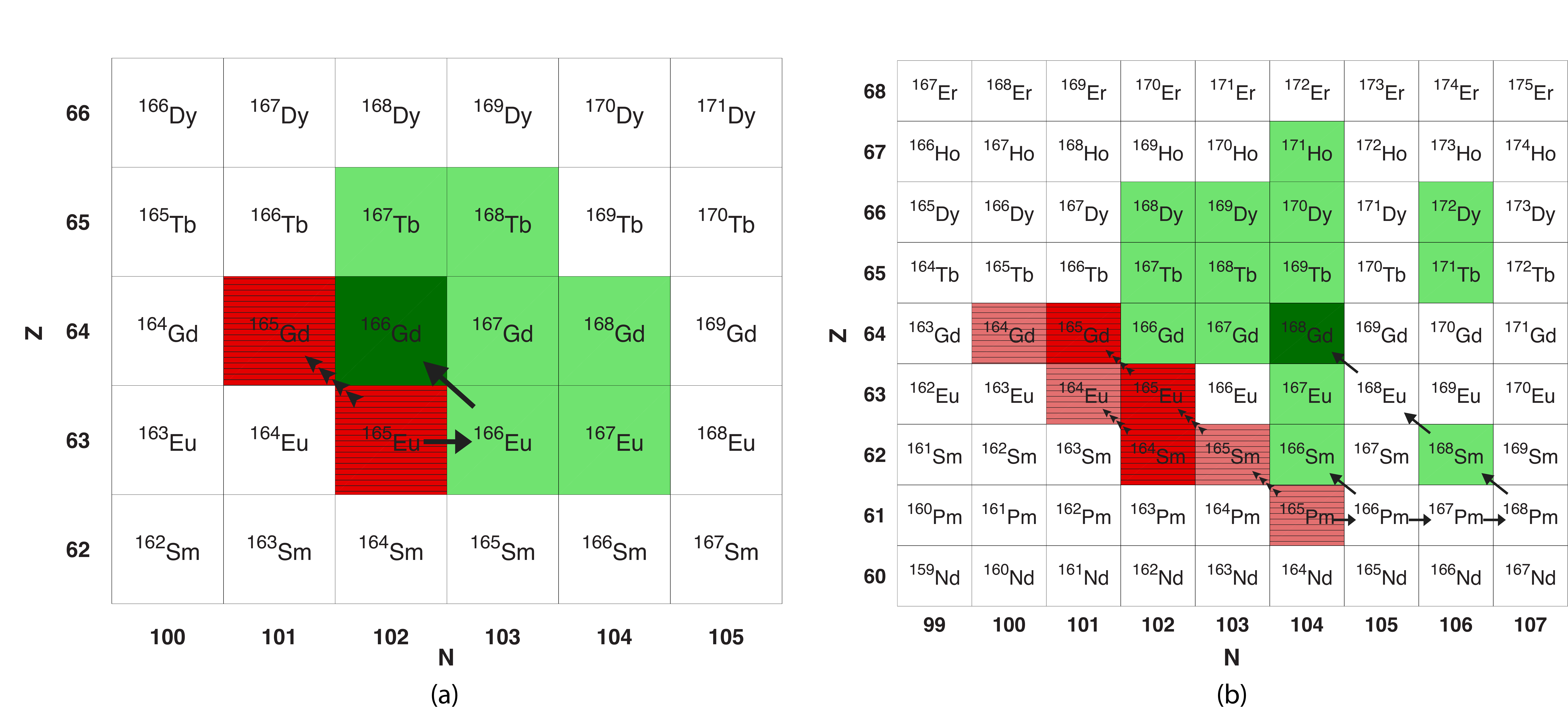}}
      \caption{\label{fig:path2} Shows the change in nucleosynthetic pathway when the neutron capture rate of $^{165}\text{Eu}$ (left panel) and $^{165}\text{Pm}$ (right panel) are modified by a factor of $K=10$. In the left panel we study the $^{165}\text{Eu}$ neutron capture effect under the hot trajectory. In the right panel we study the $^{165}\text{Pm}$ neutron capture effect under the cold trajectory. In both cases the baseline simulations exhibit flow dominated by $\beta$-decay. The neutron capture effect of $^{165}\text{Eu}$ changes the pathway resulting in most material being deposited in $^{166}\text{Eu}$ and $^{166}\text{Gd}$. The neutron capture effect of $^{165}\text{Pm}$ represents an extreme case where multiple new channels are opened. The markings are the same as in Figure \ref{fig:path1}.}
\end{figure*}

For each neutron capture rate study we normalize the set of $\Delta Y$'s and represent their magnitude by colors in the NZ-plane of Figures \ref{fig:path1} and \ref{fig:path2}. Darkest solid shades (online dark green) represent the largest order of magnitude positive change while light solid shades (online light green) represent the next largest positive change. Darkest hatches (online dark red) represent the largest order of magnitude negative change while light hatches (online light red) represent the next largest negative change.

Pathway changes can vary between the same nuclei under different simluations for a number of reasons. This includes variations due to astrophysical conditions, the onset of $r$-process freeze-out, availability of free neutrons, large neutron capture flow or flow saturation.

Of the ten elements shown in Figure \ref{fig:abncr} we select four of them for a study of the pathway changes. We examine Neodymium-156 in the context of pathway changes due to differences in astrophysical conditions. We then consider pathway changes of two different nuclei Europium-165 and Promethium-165 to highlight the importance of large neutron capture flow of surrounding nuclei.

In Figure \ref{fig:path1} we see the changes in nucleosynthetic pathways of Neodymium-156, which are slightly greater in the classical senario than in the cold senario. In both cases, before changing the neutron capture rate $^{156}\text{Nd}$ was populated from a $\beta$-decay channel from $^{156}\text{Pr}$. Because of the quick $\beta$-decay rate of $^{156}\text{Nd}$ the material continued to flow to $^{156}\text{Pm}$; following the dotted arrows.

After increasing the neutron capture rate by a factor of $K=10$, under the hot trajectory (left panel), the neutron capture channel is enhanced. The flow of material now travels through the neutron capture channel rather than the $\beta$-channel resulting in material being deposited in new elements as highlighted in a light solid shade; follow solid arrows.

Under the cold evolution we find a similar effect. However, the neutron capture rate of $^{156}\text{Nd}$ in the cold environment baseline is half the value of the capture rate of $^{156}\text{Nd}$ in the hot environment baseline. At the time when $\dot{S}$ is maximal, less material will travel through the capture channel in the cold senario with increase in the capture rate resulting in a slightly more constrained pathway. We see this in Figure \ref{fig:path1} by comparing panel (a) and (b). In both environments $^{156}\text{Nd}$ reaches saturation around $K=10$.

In Figure \ref{fig:path2} the nucleosynthetic pathway changes of Europium-165 are displayed for the hot evolution (left panel) and the pathway changes of Promethium-165 for the cold evolution (right panel) under a capture rate change of $K=10$. These two nuclei's capture rates have similar effects on the abundance pattern, $F\approx400$, but the pathway changes for $^{165}\text{Pm}$ extend through nine units of atomic mass while the pathway changes of $^{165}\text{Eu}$ extend through only four. This discrepancy arises from the differences in the flow through the neutron capture channel of the surrounding nuclei.

In the case of $^{165}\text{Eu}$ the capture rate change is important during relatively late times. At this point in the hot simulation the temperature has fallen drastically and free neutrons are relatively scarce. The unchanged neutron capture channels of nuclei surrouding $^{165}\text{Eu}$ have difficulty competing with the increasing $\beta$-decay flow in the region. The new pathway is thus limited from branching out resulting in most material being shifted to $^{166}\text{Gd}$.

For $^{165}\text{Pm}$ the capture rate change is important farther from stability (at earlier times) in the cold evolution and it also does not have to compete with photo-dissociation flows. Note that $^{165}\text{Pm}$ has no effect on the abundance pattern under the hot evolution, see bottom panel of Figure \ref{fig:fgrid}. In addition, $^{165}\text{Pm}$ is an even-N nucleus so that it is far from saturation compared to odd-N nuclei populated at early times. At this point in the cold simulation the unchanged neutron capture rates of the surrounding nuclei can still compete with $\beta$-decay rates in the region resulting in the formation of many secondary pathways beyond $^{166}\text{Pm}$.

\section{Neutron Capture Rate Decreases \& Nuclear Physics Uncertainties}\label{sec:npu}
For completeness, we also conducted neutron capture rate studies where the rates were decreased by factors of $5$, $10$, $50$ and $100$. The same general analysis presented above can be applied to neutron capture rate decreases. The major difference arises when saturated nuclei (in the baseline simulation) are moved further from saturation by decreasing neutron capture rate, reducing the second term in equation \ref{eqn:sat}.

\begin{figure*}[htp]
      \includegraphics[width=175mm,height=200mm]{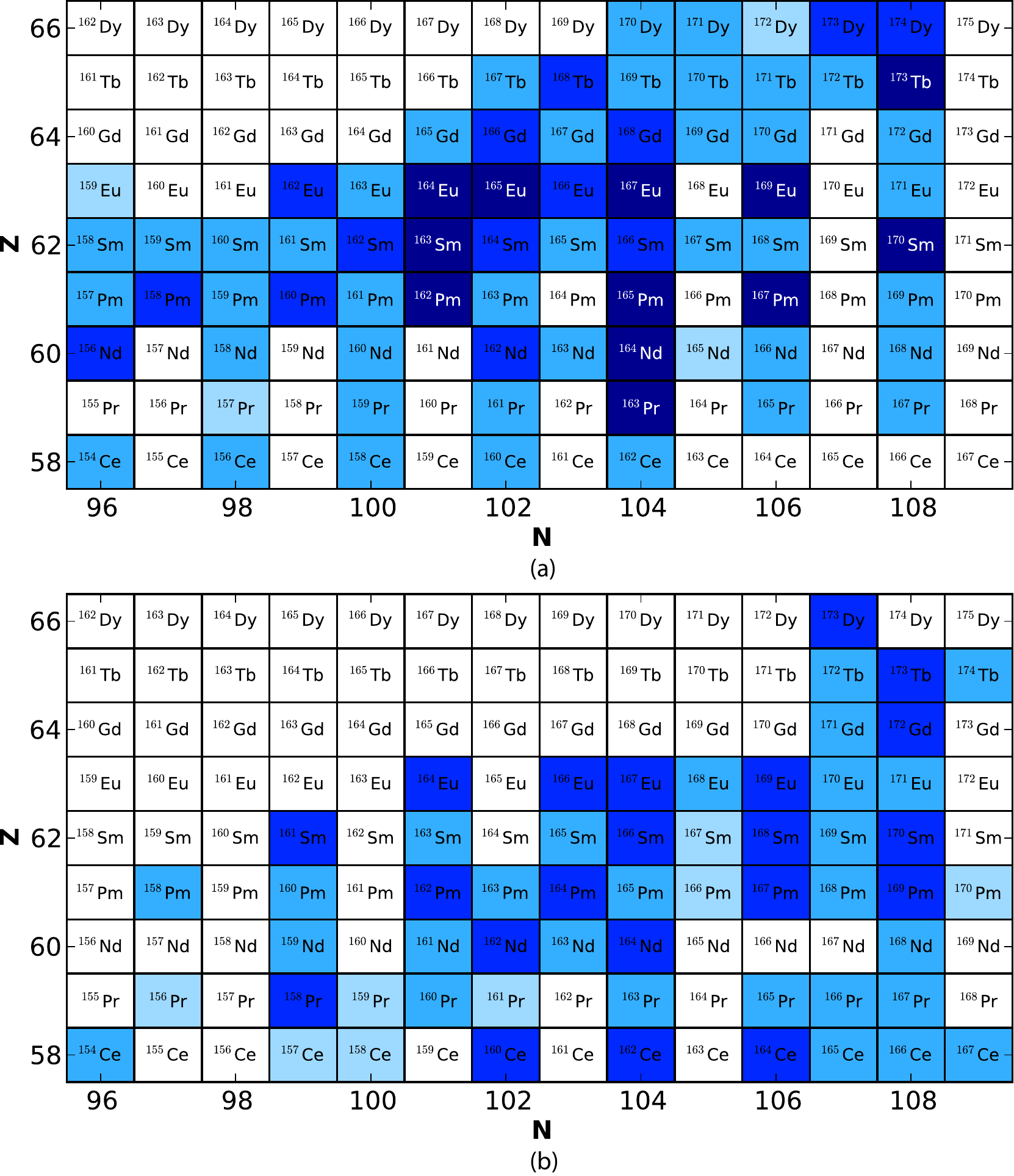}
      \caption{\label{fig:rgrid} Identifies all nuclei in a combined data set whose neutron capture rates can significantly impact the final abundance pattern when rates are (a) increased and (b) decreased. The data set includes calculations using a combination of nuclear models (FRDM,ETFSI,HFB17,DZ) and conditions (hot and cold). Shaded nuclei have $F_{max}\approx200$ or more abundance change for a neutron capture rate increase or decrease of $K=5$ (darkest shaded squares), $K=10$ (medium shaded squares), $K=50$ (light shaded squares) and $K=100$ (lightest shaded squares). For each nuclei the maximal $F$ was chosen among the data sets. Nuclei shaded white never produce a significant effect ($F\gtrsim200$) under any data set for any rate change, $K$.}
\end{figure*}

To summarize the effects of nuclei whose neutron capture rates can significantly impact abundance patterns we generated a combined data set which includes four nuclear models (FRDM,ETFSI,HFB17,DZ) and two astrophysical conditions (hot,cold) studying both increases and decreases by factors of $5$, $10$, $50$ and $100$. We separated the data based on whether the rate was increased or decreased. We then selected the maximal $F$ for each nuclei across these studies, the results of which are shown in Figure \ref{fig:rgrid}.

Panel (a) shows those nuclei whose neutron capture rates can significantly impact final abundances when the rates are increased. The lengthy freeze-out phase of cold evolutions contributes to most of the shaded nuclei in the bottom right portion of the panel. Here, Even-N nuclei are favored across different nuclear data sets due to flow saturation in odd-N nuclei. The neutron capture effect of nuclei near the rare earth peak, with $N=102$, $N=104$ and $N=106$ and from Praseodymium up to Europium is exhausted for changes in neutron capture rates of only $K=5$. Capture rate changes above $K=5$ do not matter for these nuclei as their neutron capture channel becomes exhausted due to saturation.

Panel (b) shows those nuclei whose neutron capture rates can significantly impact final abundances when the rates are decreased. We find that decreases in neutron capture rates in the rare earth region tend to have smaller effects than increases. Because neutron capture rate decreases restrict the flow through the neutron capture channel we find that larger neutron flow, $L\gtrsim.8$, is required to produce significant change to the final abundance pattern. Thus, the distribution of influential neutron capture rates is shift to more neutron rich nuclei.

\section{Conclusions}\label{sec:conclusions}
We have demonstrated the importance of understanding individual neutron capture rates in the rare earth region of the $r$-process abundance pattern and shown that the distribution of influential nuclei can be elucidated by the concepts of large nuclear flow and flow saturation. These concepts are applicable across a variety of astrophysical conditions and nuclear models. An influential neutron capture rate leads to a `neutron capture effect' where a change is effected in the abundance of nearby higher $A$ nuclei.

Many nuclei show significant leverage on the final abundances with small neutron capture rate change (by a factor of $K=5$), as shown in Figure \ref{fig:rgrid}. The distribution of important rare earth neutron capture rates in the NZ-plane occurs in a narrow band approximately $10$ to $20$ neutrons from stability. The overall distribution is remarkably similar across a variety of freeze-out conditions and differing input nuclear physics, e.g Figure \ref{fig:rgrid} and is in agreement with the prediction of the location of important neutron capture rates in the rare earth region based on formation arguments \cite{Mumpower2011}.

To understand the magnitude of the neutron capture effect we introduced two concepts: (1) large nuclear flow and (2) flow saturation. Large nuclear flow means significant material transportation in a given reaction channel. The requirement of large nuclear flow in the neutron capture channel tends to rule out nuclei that are less than $10$ neutrons from stability in the rare earth region as these nuclei are populated primarily via $\beta$-decay. Flow saturation occurs when the sum of material flowing through input channels matches the flow of a single output channel. In the case where the output channel is neutron capture, the channel is `saturated' when it is matched by photo-dissociation and $\beta$-decay in-flows. Changes to a neutron capture rate under saturation have no effect on the flow of material because the output channel is limited by the in-flowing channels. In like manner, changes in capture rates by very large factors are also forbidden because the input channels become exhausted.

Flow saturation is useful in understanding the details of the pattern of influential nuclei. For instance, under hot freeze-out conditions, photo-dissociation rates are large so that the net flow to the right in the NZ-plane can be limited by leftward flowing material in the photo-dissociation channel. This limits the neutron capture effect for nuclei far from stability either in \nggn \ equilibrium or for those nuclei just coming out of equilibrium because the channel is saturated. Under cold freeze-out conditions, odd-N nuclei are closest to saturation since they tend to have faster neutron capture rates and $\beta$-decay rates. Thus, large neutron capture effects are generally found in even-N nuclei.

The concepts of large nuclear flow and flow saturation are general concepts which are applicable beyond the scope of individual neutron capture studies. We expect that the overall distribution of influential nuclei would still exhibit the same qualitative behavior, if for instance, one chose to study groups of neutron capture rates.

In order to disentangle the information contained in the abundances found in nature, detailed knowledge of nuclear physics including masses, $\beta$-decay rates and neutron capture rates must be known. The concepts of large nuclear flow and flow saturation introduced here are essential concepts for studying any reaction rates relevant to the $r$-process.

\section{Acknowledgments}
We thank Raph Hix for providing the charged particle network and up-to-date reaction libraries. 
We thank T. Rauscher for valuable discussions regarding neutron capture rates at low temperature.
We thank North Carolina State University for providing the high performance computational resources necessary for this project. 
This work was supported in part by U.S. DOE Grant No. DE-FG02-02ER41216, DE-SC0004786, and DE-FG02-05ER41398.

\bibliographystyle{unsrt}
\bibliography{rep-ncr-refs}

\end{document}